\documentclass[journal]{IEEEtran}
\IEEEoverridecommandlockouts
\ifCLASSINFOpdf
\else
\fi

\usepackage{amssymb}
\usepackage{amsmath}
\usepackage{amsthm}
\usepackage{graphicx}
\usepackage{algorithm}
\usepackage{algorithmic}
\usepackage{booktabs}
\usepackage{cite}
\usepackage{bm}
\usepackage{float}
\usepackage{multirow}
\usepackage{color}
\usepackage{subfigure}
\usepackage{caption}
\usepackage{epstopdf}
\usepackage{longtable}
\usepackage{diagbox}
\usepackage{changepage}
\usepackage{comment}
\usepackage{cases}
\usepackage[short]{newoptidef}
\usepackage{makecell}
\usepackage{mathabx}
\theoremstyle{definition}

\theoremstyle{remark}

\theoremstyle{plain}

\newcommand{\tmop}[1]{\ensuremath{\operatorname{#1}}}

\begin{document}
		%
		\title{Hybrid Reinforcement Learning-based Sustainable Multi-User Computation Offloading for Mobile Edge-Quantum Computing
		\author{Minrui~Xu, Dusit~Niyato,~\emph{Fellow, IEEE}, Jiawen~Kang, Zehui~Xiong, Mingzhe~Chen, \\Dong In Kim,~\emph{Fellow, IEEE}, and Xuemin (Sherman) Shen,~\emph{Fellow, IEEE}}
  \thanks{Part of this article was presented at the IEEE International Conference on Communications 2023 \cite{xu2023learning}.}
    \thanks{M.~Xu and D.~Niyato are with the College of Computing and Data Science, Nanyang Technological University, Singapore (e-mail: minrui001@e.ntu.edu.sg; dniyato@ntu.edu.sg). J.~Kang is with the School of Automation, Guangdong University of Technology, China (e-mail: kavinkang@gdut.edu.cn). Z.~Xiong is with the Pillar of Information Systems Technology and Design, Singapore University of Technology and Design, Singapore 487372, Singapore (e-mail: zehui\_xiong@sutd.edu.sg). M.~Chen is with the Department of Electrical and Computer Engineering and Institute for Data Science and Computing, University of Miami, Coral Gables, FL, 33146 USA (e-mail: mingzhe.chen@miami.edu). D.~I.~Kim is with the Department of Electrical and Computer Engineering, Sungkyunkwan University, Suwon 16419, South Korea (email: dongin@skku.edu). X.~Shen is with the Department of Electrical and Computer Engineering, University of Waterloo, Waterloo, ON N2L 3G1, Canada (e-mail: sshen@uwaterloo.ca).}
 }
		\maketitle
		\begin{abstract}
			
  Exploiting quantum computing at the mobile edge holds immense potential for facilitating large-scale network design, processing multimodal data, optimizing resource management, and enhancing network security. In this paper, we propose a pioneering paradigm of mobile edge quantum computing (MEQC) that integrates quantum computing capabilities into classical edge computing servers that are proximate to mobile devices. To conceptualize the MEQC, we first design an MEQC system, where mobile devices can offload classical and quantum computation tasks to edge servers equipped with classical and quantum computers. We then formulate the hybrid classical-quantum computation offloading problem whose goal is to minimize system cost in terms of latency and energy consumption. To solve the offloading problem efficiently, we propose a hybrid discrete-continuous multi-agent reinforcement learning algorithm to learn long-term sustainable offloading and partitioning strategies. Finally, numerical results demonstrate that the proposed algorithm can reduce the MEQC system cost by up to 30\% compared to existing baselines.
		\end{abstract}
		

		%
		\IEEEpeerreviewmaketitle

		\section{Introduction}
		
Quantum computing and communication are envisioned as strategic technologies across academic and industrial sectors, since they will introduce significant advantages in extant technological fields, including artificial intelligence (AI)~\cite{biamonte2017quantum}, security~\cite{broadbent2016quantum}, and finance~\cite{herman2022survey}. 
For instance, the Quantum Internet can leverage quantum optimization algorithms, which have the potential to provide faster and more efficient processing of large amounts of data~\cite{10123997, rabbie2022designing, wang2022quantum}. Moreover, with quantum computing, resource management in the Quantum Internet can be optimized by improving the allocation and utilization of resources at the edge of networks, leading to more efficient use of available resources. Additionally, quantum cryptography can enhance security in mobile edge networks by providing a higher level of protection against cyberattacks and security threats~\cite{herman2022survey}. Finally, by enabling faster and more efficient data processing, quantum computing can help streamline data management in mobile edge networks, making it easier to manage and analyze large-scale data in real-time.
  Despite these promising applications, executing such quantum computation tasks necessitates scalable quantum computers endowed with an approximate capacity of 10$^6$ qubits.
		
Diverging from classical computing~\cite{chen2015efficient, yuan2022digital}, quantum computing faces unique challenges as the number of qubits, quantum gates, and measurement operations in scalable quantum computers~\cite{caleffi2022distributed}. Among these challenges, the most significant one is the inherent qubit noise that might compromise the fidelity of quantum computing~\cite{resch2021benchmarking}. To achieve fault-tolerant quantum computing, scalable quantum computers can operate quantum processing units (QPUs) with advanced cryogenic components and fault-tolerant schemes~\cite{martin2022energy}. On the one hand, quantum computers work at extremely low temperatures to cool quantum devices to a low-entropy state. On the other hand, quantum noise is combated through fault-tolerant schemes, including concatenated codes, surface codes, and bosonic qubits~\cite{resch2021benchmarking}. For instance, using error-correction codes allows quantum information to be maintained across several physical qubits, constituting one logical qubit. Overall, the tremendous computing and energy advantages can be achieved only with fault-tolerant quantum computing operating at cryogenic conditions.
		

As scalable quantum computers meet the requisite scale and quality parameters, mobile edge-quantum computing (MEQC)~\cite{10778612}, coupled with the remote accessibility offered by edge servers, may extend the reach of quantum advantages into mobile edge networks. By offloading computation tasks to quantum computers in edge servers, users can gain significant benefits from cloud/edge quantum computer providers, such as Amazon Braket~\cite{aws}, IBM Quantum\cite{ibm}, and Azure Quantum\cite{azure}. This extends the potential of discovering innovative applications for the Quantum Internet and tackling existing issues, thus driving the practical implementation of scalable quantum computers.
		MEQC can provide users with a range of potentially lethal applications, such as quantum ray tracing~\cite{santos2022towards}, by significantly increasing the appeal of quantum computing to mobile consumers. Unlike classical edge computing~\cite{chen2015efficient}, MEQC has unique differences in the measurement of computing power and energy consumption. First, quantum computing uses the superposition, interference, and entanglement of qubits to accelerate the execution of computation tasks. 
        Second, quantum computers operate in an extremely low-temperature environment. Therefore, compared to conventional computers, most of the energy consumption of quantum computers is used to maintain the ultra-low temperature container. Third, for the reliability of execution results, quantum computers choose appropriate error correction codes and the concatenation degree of error correction according to their latency and energy constraints. Nonetheless, the integration of quantum computers into mobile edge networks to execute quantum computation tasks introduces complexities that go beyond those that are inherent in classical MEC. Finally, to determine offloading strategies for mobile devices to perform hybrid classical-quantum computation tasks in edge servers efficiently is still challenging.
		\begin{figure}[t]
    \centering
    \includegraphics[width=1\linewidth]{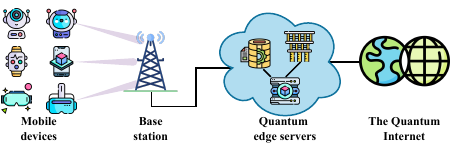}
    \caption{An illustration of mobile edge-quantum computing in the Quantum Internet.}
    \label{fig:system}
\end{figure}

		In the Quantum Internet, the problem of determining sustainable strategies in the hybrid classical-quantum computation offloading problem involves continuous and discrete optimization variables in both classical and quantum computing. Therefore, deciding on how much computation and which server to offload is a complex mixed-integer programming problem with the non-convex and non-linear objective of minimizing computational
cost in terms of latency and energy consumption. Fortunately, deep reinforcement learning (RL)~\cite{neunert2020continuous, chen2020joint,chen2021distributed} is used by mobile devices to learn a sustainable offloading and partitioning strategy without prior knowledge of the computation tasks, edge server status, and local environment. To minimize long-term system cost, deep RL can let mobile devices act as agents to learn and obtain sustainable offloading and partitioning strategies under dynamically changing quantum noise and the uncertain requests of computation tasks. By leveraging hybrid discrete-continuous policies to interact with the MEQC environment, mobile devices can learn to determine the discrete offloading and the continuous partitioning decisions for classical-quantum computation tasks to maximize the sustainability of the MEQC system. Finally, we leverage variational quantum circuits (VQCs) to parameterize the actor-critic networks of agents, which can accelerate the convergence process of learning-based algorithms without the loss of performance, and thus achieving a higher sustainability.
		
		The main contributions can be summarized as follows.
		\begin{itemize}
		    \item We propose a novel paradigm of mobile edge-quantum computing in the Quantum Internet that brings quantum advantages to mobile edge networks, and design a mobile edge-quantum computing system where mobile devices can flexibly offload hybrid classical-quantum computation tasks to edge servers.
		    \item Based on the MEQC system model, we formulate a hybrid classical-quantum computation offloading problem whose goal is to minimize non-convex and nonlinear system costs in terms of latency and energy consumption. As the quantum computing system has dynamic state space and large-scale action space, the problem is hard to tackle by conventional optimization methods.
		    \item To improve the sustainability in MEQC, we formulate the hybrid classical-quantum computation offloading problem as a partially observable Markov decision process (POMDP), which reduces the complexity of the learning-based algorithms. To reduce training resources in deep RL, we also propose a hybrid RL-based algorithm to learn the sustainable offloading and partitioning strategy with quantum neural networks.
      \item The experimental results demonstrate that the proposed hybrid discrete-continuous multi-agent RL algorithm can converge to a sustainable offloading and partitioning strategy, which can reduce the system cost by at least 30\% compared with other baseline algorithms. In addition, the proposed algorithm can accelerate the convergence of agents and improve the sustainability of the proposed learning-based algorithm.
		\end{itemize}
		
		
The rest of the paper is organized as follows. We first discuss the related works in Section II. Furthermore, we present the system model of MEQC in Section III and propose the hybrid multi-agent RL-based algorithms in Section IV. We present the simulation results in Section V and conclude in Section VI.
\section{Related Works}

\subsection{Quantum Advantage in Mobile Edge Networks}

Quantum advantage refers to the ability of quantum computers to solve certain problems faster than classical computers~\cite{arute2019quantum}. In mobile edge computing, edge servers equipped with quantum computers can accelerate the computing processes of IoT tasks such as solving large-scale optimization problems, searching large databases, and simulating quantum systems, compared with classical computing.

In the Noisy Intermediate Scale Quantum (NISQ) era~\cite{bharti2021noisy}, quantum computers with tens or hundreds of qubits are being developed and deployed in mobile edge networks to perform quantum computation tasks. For example, Wang \textit{et al.} in~\cite{wang2022quantum} discuss the advantages of quantum communication and computation in 6G networks, including radio access networks, non-terrestrial networks, edge networks, edge data centers, blockchain, and wireless artificial intelligence. In particular, Zaman \textit{et al.} in~\cite{zaman2023quantum} explore the potential of quantum intelligence, which exploits quantum acceleration to meet the stringent requirements for ultra-reliable and low-latency communications (URLLC) in 6G networks. For optimization problems of NP-hard URLLC tasks, they demonstrated quantum algorithms for tackling task relocation and accelerating machine learning in wireless networks.

To validate the effectiveness of these NISQ algorithms, both small-scale QPUs with tens of qubits and simulated quantum computing via CPUs/GPUs can be leveraged~\cite{willsch2022gpu}. PennyLane~\cite{bergholm2018pennylane} provides a unified architecture for near-term quantum computing devices. In Pennylane, the designed quantum algorithms can be tested in publicly accessible devices provided by Xanadu Cloud, Amazon Braket, and IBM Quantum. TensorCircuit~\cite{zhang2023tensorcircuit} is an open-source quantum circuit simulator with a 600-qubit capacity, which supports automatic differentiation, just-in-time compilation, vectorized parallelism, and hardware acceleration, designed for speed, flexibility, and code efficiency. 

In the fault-tolerant quantum computing era, scalable quantum computers can perform complex computation tasks with high computations reliably through quantum error correction techniques~\cite{fellous2022optimizing}. However, optimizing the resource efficiency of scalable quantum computers is complicated as error correction for mitigating the effects of noise and decoherence in quantum systems requires additional qubits and operations.


\subsection{Mobile Edge-Quantum Computing and Quantum Computation Offloading}

MEQC is a pioneering paradigm of mobile edge networks in the quantum computing era that deploys quantum computers or simulated quantum computing on edge servers and even mobile devices to bring quantum advantage to the edge. For instance, Leymann \textit{et al.} in~\cite{leymann2020quantum} discuss the commercial availability of quantum computers and their accessibility through cloud-based services. In demonstrating practical applications of quantum cloud computing, they conceive a collaborative quantum application platform, which leverages quantum machine learning for a multitude of use cases, encompassing fields such as digital humanities.
Moreover, Passian \textit{et al.} in~\cite{passian2023concept} introduce the concept of an edge quantum computing simulator, a platform conceived for the design of the next generation of edge computing applications. To enable mobile devices to engage with quantum edge applications, the authors suggest the development of initial quantum edge simulators to provide a comprehensive framework, wherein quantum processors and algorithms can proficiently manage noisy data, data processing, error correction, optimization, and communication. 

Within the edge-cloud continuum, Furutanpey \textit{et al.}~in\cite{furutanpey2023architectural} underscore the possibilities of incorporating QPUs, alongside presenting an architectural vision for edge-cloud quantum computing. They introduce a distributed inference engine armed with hybrid classical-quantum neural networks (QNNs) to assist system designers in catering to applications with intricate requirements that engender the highest degree of heterogeneity. Analogous to classical computation offloading, quantum computation offloading entails the transfer of quantum computation tasks from mobile devices to servers fortified with quantum computers or quantum computing simulators for execution.
Speer~\textit{et al.} in~\cite{speer2021program} explore the viability of quantum computation offloading through the use of program equivalence checking for the automatic identification of code compatible with quantum offloading.


\subsection{Deep RL and Quantum Computing in Computation Offloading}

Deep RL and Quantum Computing can be leveraged in computation offloading to augment decision-making processes and enhance computational efficiency. In wireless-powered mobile-edge computing networks, Huang~\textit{et al.} in~\cite{huang2019deep} propose a Deep RL-based Online Offloading (DROO) framework. This approach utilizes a deep neural network as a scalable solution that gleans binary offloading decisions from experiential learning. Focused on delay-oriented task offloading in Self-organizing Architecture based on Generalized Information Network (SAGIN), Zhou~\textit{et al.}~in~\cite{zhou2020deep} introduce a unique deep risk-sensitive RL algorithm. This tool aims to minimize partial offloading and computing delay of all tasks given the constraints of Unmanned Aerial Vehicle energy capacity. In the context of offloading games in edge computing, Zhan \textit{et al.} in~\cite{zhan2020deep} propose a decentralized offloading algorithm based on deep RL. In this scenario, agents with incomplete information can learn offloading decisions by interacting with the environment. 

Given that real-world mobile edge computing (MEC) systems tend to comprise a vast number of users, servers, and hybrid discrete-continuous decisions, Ho~\textit{et al.}~in\cite{ho2020joint} employ a hybrid deep RL-based algorithm for joint server selection, partial offloading, and handover decisions within a multi-access edge wireless network. Nonetheless, none of the aforementioned works amalgamate the benefits of RL and quantum computing in making joint offloading and partitioning decisions within MEC.  Beyond deep RL, Dong~\textit{et al.}~\cite{dong2022quantum} utilize the quantum advantage in MEC by employing a quantum particle swarm optimization-based approach. This method is designed to solve the optimization formulation defined for multi-user multi-server task offloading.

To the best of our knowledge, we are the first to propose the MEQC and leverage multi-agent RL algorithms to optimize the non-convex and non-linear hybrid classic-quantum computation offloading problems with mixed-integer variables.

\section{System Model and Problem Formulation}
  We consider the MEQC system that consists of a set $\mathcal{U}$ of $U$ mobile devices and a set $\mathcal{E}$ of $E$ quantum edge servers. We assume that each user $u$ requests a computation task $\mathcal{T}_u^\emph{C} \triangleq (s_u,n_u)$ that can be executed partially by local CPUs and one edge server that has both CPUs and QPUs. Here, $s_u$ represents the data size of each raw computation task and $n_u$ represents the required CPU cycles per data size to accomplish the computation task. Next, we first introduce the delay and energy that each device processes a partial computation task with its local CPUs.

  \subsection{Offloading Classical-Quantum Computation Tasks for Mobile Edge-Quantum Computing}
\begin{figure}[t]
	\centering
	\includegraphics[width=1\linewidth]{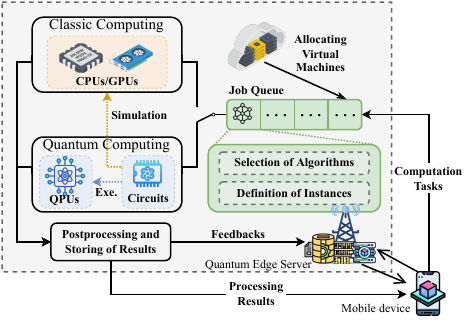}
	\caption{The workflow of hybrid classical-quantum computation offloading for mobile edge-quantum computing in the Quantum Internet.}
	\label{fig:workflow}
\end{figure}
Classical MEC refers to a system that enables mobile devices to offload computationally-intensive applications to proximate edge servers for remote execution via wireless connections. 
However, the integration of quantum computers into mobile edge networks for the execution of quantum computation tasks introduces complexities beyond those inherent in classical MEC~\cite{sim2018framework}. 

As shown in Fig.~\ref{fig:workflow}, additional processes, such as the selection of algorithms, instance definition, and quantum circuit compilation, must be taken into account when operating quantum computing applications in MEQC, users initially select a quantum algorithm that aligns with their problem. This might be a problem-specific algorithm or a more versatile one, such as the Variational Quantum Eigensolver (VQE) or Quantum Approximate Optimization Algorithm (QAOA). Following the algorithm selection, a specific instance of the algorithm is defined, typically articulated in terms of quantum circuits that compute the values of the objective functions for the chosen algorithm. Subsequently, these abstract quantum circuits undergo compilation based on the specifications of the quantum device in use. This process necessitates mapping the abstract qubits onto physical qubits within the device and decomposing the abstract gates into the device's native gates. Upon receiving tasks from mobile devices, edge servers assign virtual machines to handle the requests and place them into job queues for execution. Rather than transmitting the compiled quantum circuits to centralized cloud-based quantum computers, these are directed to quantum computers on edge servers. Since these quantum computers are physically closer to the data source, computational latency is reduced. The compiled quantum circuits are then executed on the quantum computers of the edge servers. The device executing quantum algorithms may be QPUs or a reliable simulator running on CPUs/GPUs, contingent on the capabilities of the quantum computing infrastructure. The quantum computation results, encompassing a collection of measurement results, undergo post-processing at the edge. The values of the algorithm's objective functions are computed, and any requisite error corrections are applied. Both the processed data and the raw data can be stored at the edge for further analysis or returned to the data source. Optimization of algorithm parameters is conducted by a classical computer. After updating parameters, the modified circuits are recompiled and re-executed on the edge quantum device. This sequence repeats until the algorithm converges toward a solution. In a manner similar to the MEC workflow, the results of the performed tasks are transmitted back to mobile devices, which could influence future data generation. 

  \subsection{Local Computing Model}
  To define the system cost of the MEQC system, we first introduce the latency of each user processing a partial computation task.
  We consider that the computational capacity of each user $u$ is $f_u^L$ (i.e., CPU cycles per second), and the proportion of a task that user $u$ processes locally is $\phi_u$. 
  Then, the latency of user $u$ processing a partial task $\phi_u s_u$ locally is $d_u^L(\phi_u) = \frac{\phi_u s_u n_u}{f_u^L}$. Meanwhile, the energy consumption of user $u$ processing a partial task $\phi_u s_u$ is $e^L_u(\phi_u) = \gamma \phi_u s_u n_u$, where $\gamma$ is the chip coefficient.
  The cost of user $u$ processing partial computation task $\phi_u s_u$ is
		\begin{equation}
		    c_u^L(\phi_u) = \lambda_u^\emph{D} d_u^L(\phi_u)+ \lambda_u^\emph{E} e^L_u(\phi_u),
		\end{equation}
		where $\lambda_u^\emph{D}, \lambda_u^\emph{E}\in [0, 1]$ denote the weight parameters of serving latency and energy, respectively. 
  
  \subsection{Edge Computing Model}

  Next, we introduce the energy and delay that each user offloads its computation task $\left(1-\phi_u\right)s_u$  to target edge server $a_u \in \mathcal{E}$. Here, task offloading consists of the computation task transmission phase and the computational processing phase. During the task processing phase,  each server can use either CPUs or quantum computing to process its received computation task. We assume that each server can use its quantum computing to process only one task per time slot~\cite{resch2021benchmarking}. To begin with, we first introduce the basics of quantum computing.

  \subsubsection{Basics of Quantum Computing}\label{sec:basics} 

By leveraging quantum mechanics, including entanglement and superposition, information of classical bits can be encoded into quantum bits, or qubits, which can be not only in the state $|0\rangle$ and $|1\rangle$ but also their superposition $\alpha|0\rangle+\beta|1\rangle$, where $\alpha,\beta\in\mathbb{C}$ and $|\alpha|^2 + |\beta|^2 = 1$.
The superposition of $n$ qubits can be represented by $|b\rangle^{\otimes n} = \sum_{i=0}^{2^n-1}\alpha_i|i\rangle$, where $\forall \alpha_i\in\mathbb{C}$ and $\sum_{i=0}^{2^n-1}|\alpha_i|^2 = 1$. Based on the superposition, the system can represent $N = 2^n$ states simultaneously with $n$ qubits. This provides quantum advantages to computation due to exponential quantum parallelism. 

The manipulation of qubits is achieved by quantum gates, including unitary gates and measurement gates. In detail, unitary gates implement unitary transformations of quantum states and measurement gates implement probabilistic and destructive transformations for classical information extraction from quantum states. For instance, the Pauli-X gate, often likened to the classical NOT gate, flips the state of a qubit from $|0\rangle$ to $|1\rangle$ and vice versa. Another example, the Hadamard gate, is pivotal in creating superposition—a quintessential quantum phenomenon—by mapping the $|0\rangle$ state to $\frac{1}{\sqrt{2}}(|0\rangle + |1\rangle)$ and $|1\rangle$ state to $\frac{1}{\sqrt{2}}(|0\rangle -|1\rangle)$, thereby creating an equal probability of being in either state upon measurement.

Measurement gates are used to extract classical information from quantum states through probabilistic and destructive transformations. These gates essentially ``collapse" the superposition, selecting one state with a certain probability and outputting a classical bit. If a measurement gate is applied to this qubit, it will randomly collapse to either the $|0\rangle$ or $|1\rangle$ state, giving the result of a classical bit, either 0 or 1. It is important to note that this process is destructive because once measured, the original quantum state is lost and cannot be replicated or reversed.

Building on qubits and quantum gates, various quantum circuits can be designed to implement corresponding quantum algorithms to achieve effective acceleration of classical computation tasks. Through unitary gates that navigate the complex landscape of quantum states, and measurement gates that bridge the quantum and classical worlds, quantum computing leverages the peculiar properties of quantum mechanics to perform computations that are currently beyond the reach of classical machines. Let $\mathbf{a} = \{a_1,\ldots, a_U\}$ denote the offloading decisions of mobile devices.

\subsubsection{Task Transmission}
  The latency and energy that user $u$ uses to transmit the data with size $\left(1-\phi_u\right)s_u$ are given by
    \begin{equation}\label{eq:transmissiond}
        d_u^\emph{O}(\phi_u,\boldsymbol{a}) = \frac{(1-\phi_u)s_u}{r_u(\boldsymbol{a})},
    \end{equation}
    and
    \begin{equation}
        e_u^\emph{O}(\phi_u,\boldsymbol{a}) = \frac{p_u (1-\phi_u) s_u}{r_u(\boldsymbol{a})},
    \end{equation}
  respectively. Here, $r_u(\boldsymbol{a}) = B \log_2 (1 + \frac{p_u g_{u}(a_u)}{\sigma_{a_u}^2})$ is the uplink data rate from user $u$ to server $a_u$ and $B$ denotes the bandwidth, $g_{u}(a_u)$ is the channel gain between user $u$ and edge server $a_u$, $\sigma_{a_u}$ is the AGWN at server $a_u$, and $p_u$ denotes the transmit power of user $u$.

\subsubsection{Task Processing at Edge Server} When a server receives the computation task from $u$, it needs to determine whether to use CPUs or quantum computing to process the task. Next, we first introduce the latency and energy that the server uses CPUs to process the task of each user $u$. 
\paragraph{Classic Task Processing} Let $f_u^E$ denote the computing capacity that a server uses to process the task of user $u$. 
We assume that the subscribed computing resources should always be satisfied since the Internet/Metaverse operator can invest in the large-scale edge computing infrastructure~\cite{chen2015efficient, xu2022full}
  Then, the delay of the server processing user $u$'s partial offloaded task with size $\left(1-\phi_u\right)s_u$ is
		\begin{equation}
		    d^\emph{E}_u(\phi_u) = \frac{(1-\phi_u) s_u n_u}{f^E_u}.
		\end{equation}
The energy consumption that the server uses CPUs to process user $u$'s partial offloaded task with size $\left(1-\phi_u\right)s_u$ is 
		\begin{equation}\label{eq:energycomputation}
		    e^\emph{E}_u(\phi_u) = \gamma (1-\phi_u) s_u n_u.
		\end{equation}
Given (\ref{eq:transmissiond})-(\ref{eq:energycomputation}), the total cost of a server processing user $u$'s offloaded task with size $\left(1-\phi_u\right)s_u$ is expressed by 
\begin{equation}
\begin{aligned}
    c_u^E(\phi_u,\boldsymbol{a}) = \lambda_u^\emph{D} &\left[d_u^\emph{O}(\phi_u,\boldsymbol{a})+d_u^\emph{E}(\phi_u)\right]  \\& + \lambda_u^\emph{E} \left[e_u^\emph{O}(\phi_u,\boldsymbol{a})+e_u^\emph{E}(\phi_u)\right].
\end{aligned}
\end{equation}



\paragraph{Quantum Task Processing}
  Here, we first introduce the process of using quantum computing to process a computation task. Then, we model the latency and energy that each server uses quantum computational resources for task processing. 
One quantum computer containing many physical qubits located at each edge server operates in an extremely low-temperature environment. Each edge server can transform one of its received computation tasks into a quantum circuit by quantum algorithms to perform quantum acceleration for the task. In particular, let $\mathcal{T}^\emph{Q}_u = (s_u, Q_u, D_u)$ be the quantum computation task compiled from the computation task of user $u$ where $Q_u$ is the required qubits to execute the quantum circuit, and $D_u$ is the length of the quantum circuit.

To construct scalable quantum computers, error correction schemes leverage multiple noisy qubits to constitute a single logical qubit with high fidelity.
Consequently, error correction schemes require many gate operations, and hence its energy power consumption is nearly independent of the used quantum algorithm actually having at the logical level~\cite{fellous2022optimizing}. 
Let $N_{1}$, $N_{2}$, and $N_\textrm{M}$ be the average numbers of physical one-qubit (1qb) gates, two-qubit (2qb) gates, and measurement gates, respectively, run in parallel per time step of the circuit. The cost of running quantum computers in MEQC consists of latency and energy costs. Specifically, the latency in quantum computing is mainly caused by the operation of logical gates in quantum circuits~\cite{fellous2022optimizing}, which can be calculated as
\begin{equation}
    d_u^\emph{Q}(\phi_u) = (1-\phi_u)s_u Q_u \big[\tau_{1}N_{1} + \tau_{2}N_{2} + \tau_\textrm{M}N_\textrm{M}\big],
\end{equation}
where $\tau_{1}$, $\tau_{2}$, and $\tau_\textrm{M}$ are the latency of processing 1qb gates, 2qb gates, and measurement gates, respectively.

Quantum computers are operated at a cryogenic temperature for the low-entropy state, and thus, initial states of qubits can be prepared accurately. Therefore, the energy of quantum computing is mainly caused by the cooling system used to maintain the low temperature~\cite{martin2022energy}. The energy consumption of quantum computers~\cite{fellous2022optimizing} in edge servers can be defined as
\begin{equation}
    e_u^\emph{Q}(\phi_u) = (1-\phi_u) s_u Q_u \big[P_{1}N_{1} + P_{2}N_{2} + P_\textrm{M}N_\textrm{M} + P_\textrm{Q} Q\big],
\end{equation}
where $Q$ is the number of physical qubits in one logical qubit, $P_1, P_2, P_\textrm{M}$, and $P_\textrm{Q}$ are the energy consumption of each physical 1qb gate, 2qb gate, measurement gate, and qubit, respectively. The specific equations of $P_1, P_2, P_\textrm{M}$, and $P_\textrm{Q}$ are shown in Appendix~\ref{app:energy}. 

The total cost of a server using quantum computing to process user $u$'s offloaded task in terms of running latency and energy of quantum processors is
\begin{equation}
\begin{aligned}
    c_u^\emph{Q}(\phi_u,\boldsymbol{a}) = \lambda_u^\emph{D} &\left[d_u^\emph{O}(\phi_u,\boldsymbol{a})+d_u^\emph{Q}(\phi_u)\right] \\&+ \lambda_u^\emph{E} \left[e_u^\emph{O}(\phi_u,\boldsymbol{a})+e_u^\emph{Q}(\phi_u)\right].
\end{aligned}
\end{equation}

Although ultra-low temperature environments and error correction schemes are used to improve the scalability of quantum computing, the inherent noise in quantum circuits cannot be completely eliminated~\cite{krinner2022realizing}. Therefore, the success probability is used to describe the performance of quantum circuits under nondeterministic quantum operations. As the width and depth of quantum circuits increase, the number of locations where errors can occur in quantum circuits also increases, which leads to a lower success probability of edge quantum computing~\cite{resch2021benchmarking}. By employing the common noise model discussed and error probability $\epsilon_{\textrm{err}}$ per physical gate calculated in Appendix~\ref{app:noise}, the linear approximation of the success probability is~\cite{fellous2022optimizing}
\begin{equation}
  \mathcal{M}_u(a_u) = 1 - \mathcal{N}^L_u \epsilon_{\tmop{thr}} (\epsilon_{\tmop{err}}
  / \epsilon_{\tmop{thr}})^{2^{k_{a_u}}},
\end{equation}
where $\mathcal{N}^L_u = Q^L_u \times D^L_u$ denotes the number of locations where
logical errors can happen, $\epsilon_{\tmop{thr}}$ is the threshold for error
correction, $\epsilon_{\tmop{err}}$ is the errors per physical gate, and $k_{a_u}$ is the error correction's concatenation level at edge server $a_u$.

Let $Q_u^E$ denote quantum computing capacity (i.e., number of logical qubits) at edge servers subscribed by user $u$ from the Internet/Metaverse operator. Therefore, we can have an indicator function $I_u(\boldsymbol{a})$ with $I_u(\boldsymbol{a})=1$ means that the task can be executed by quantum computers at $a_u$ where $I_u(\boldsymbol{a})=0$ otherwise, which can be expressed as 
\begin{equation}
    I_u(\boldsymbol{a}) = \begin{cases}
1, & \text{ if } Q_u \leq Q_u^E \quad \&\quad \mathcal{M}_u(a_u) \geq 2/3, \\ 
0, & \text{ otherwise.}
\end{cases}
\end{equation}
Here, the threshold success probability of quantum circuits is set to 2/3~\cite{fellous2022optimizing}, which is a classical choice for a single run of the quantum circuit.
  \subsection{Problem Formulation}
    Let $\boldsymbol{\phi} = \{\phi_1,\ldots, \phi_U\}$ denote the partitioning decisions of mobile devices. Based on the above local computing model and edge computing model, the total offloading and execution cost in MEQC can be calculated as
  \begin{equation}
  \begin{aligned}
      C(\boldsymbol{a}, \boldsymbol{\phi}) = \sum_{u\in\mathcal{U}} \bigg[c_u^L(\phi_u)  &+ (1-I_u(\boldsymbol{a})) c_u^\emph{E}(\boldsymbol{a}, \phi_u)  \\&+ I_u(\boldsymbol{a}) c_u^\emph{Q}(\boldsymbol{a}, \phi_u)\bigg].
  \end{aligned}\label{eq:cost}
  \end{equation}
    Given Eq.~\eqref{eq:cost}, the problem of minimizing the total cost of task offloading in MEQC can be formulated as
\begin{mini!}|s|[2]<b>
  {\boldsymbol{a}, \boldsymbol{\phi}}{C(\boldsymbol{a}, \boldsymbol{\phi}),}{\label{obj}}{}
  \addConstraint{\sum_{u\in \mathcal{U}} \boldsymbol{1}_{\{a_u=e\}}I_u(\boldsymbol{a})}{\leq 1,}{\quad e\in \mathcal{E} \label{con0}}{}
  \addConstraint{a_u}{\in \{1,\ldots,E\},}{ \quad \forall u \in \mathcal{U}\label{con1}}{}
  \addConstraint{\phi_{u}}{\in [0, 1],}{\quad \forall u \in \mathcal{U}.\label{con2}}{}
    \end{mini!}
    The constraint in Eq. \eqref{con0} means that each edge server can only perform one quantum computation task from MEC. Moreover, the offloading constraint in Eq. \eqref{con1} means that each user can only select one of the edge servers to offload. Finally, the constraint in Eq. \eqref{con2} represents that the partitioning decision is between 0 and 1. 

    In this optimization problem, the objective structure in Eq.~\eqref{obj} is non-convex and nonlinear and the decision variables are mixed of integer values and continuous values. The state space of the mobile edge quantum computing system is dynamic and includes qubit fidelity and user requests. The action space for offloading and partitioning decisions is a mixture of discrete and continuous variables on a large scale. It is difficult to solve the optimization problem using conventional optimization methods, such as the Alternating Direction Method of Multipliers (ADMM) or Block Successive Upper-bound Minimization (BSUM). While these techniques can solve the formulated problem in any time window, they cannot guarantee a globally sustainable solution over time. Therefore, in the following section, we turn to advanced deep RL techniques to solve these problems. These techniques can effectively and efficiently determine the sustainable offloading and partitioning strategies for the hybrid problem of offloading classical and quantum computations by interacting with and learning from the environment.

\subsection{Quantum Ray Tracing via Quantum Cloud Computing}

Ray tracing is a technique used in computer graphics to generate realistic images by simulating the path of light as it interacts with objects in a scene for multimedia applications~\cite{xu2024generative}. It is a core component of most rendering techniques. In the Internet and Metaverse, ray tracing is used for creating immersive virtual environments, video games, and movies. It is also used in architectural visualization, product design, and engineering simulations.


In the quantum ray tracing algorithm proposed in~\cite{santos2022towards}, the depth range is set for the possible intersections. The depth in rendering algorithms is regarded as the distance from the intersection point to the origin of the ray. With the objective of minimum depth, the ray tracing algorithm attempts to find the primitive that intersects the closest point to the origin of the ray. Therefore, the depth range is assigned according to the parameter of depth and to the near and far fields of the ray. According to the current ray, the scene, and the depth range, the quantum ray tracing algorithm then compiles the $\hat{R}_r$ operator's quantum circuit. To estimate which primitives among a collection of $P$ primitives, $P=2^{pb}$, $pb\in \mathbb{N}$, are intersected by ray $r$, the ray tracing intersection operator $\hat{R}_r$ uses two quantum registers, i.e., the primitive register $p_{reg}$ and the indication register $i_{reg}$. The primitive register $p_{reg}$ listing all $P$ primitives is initially prepared as a uniform superposition. In addition, the indication register $i_{reg}$ is then changed to $|1\rangle$ if the $r$ is intersected with the primitive $p$ in the superposition:
\begin{equation}
	\hat{R}_{r} |0\rangle^{\otimes p b}|0\rangle \mapsto \frac{1}{\sqrt{P}} \sum_{p=0}^{P-1}\big[|p\rangle\left(\left(1-i_{r}(p)\right)|0\rangle+i_{r}(p)|1\rangle\right)\big].
\end{equation}

Using the Hadamard gates with $pb$ qubits denoted as $\hat{\mathcal{H}}_{pb}$, the superposition on $p_{reg}$ is prepared. The operator $\hat{Int}_r$ implements, for all primitives $p$ indexed by $p_{reg}$, the intersecting function $i_r(p)$ can be defined by
\begin{equation}
	i_{r}(p)=\begin{cases}
		1, & \text { if } p \text { is intersected with the ray } r, \\
		0, & \text { otherwise},
	\end{cases}
\end{equation}
and $\widehat{R}_{r}=\widehat{I n t}_{r}\left(\widehat{\mathcal{H}}_{pb} \otimes \widehat{\mathcal{I}}_{1}\right)$, where $\widehat{\mathcal{I}}_{1}$ represents the identity operator adopted to the qubit in $i_{reg}$.
Subsequently, to search for a feasible intersection within the possible range of depth, $\hat{R}_r$ as the oracle calls the quantum search algorithm~\cite{herman2022survey}. 
At termination, the quantum ray tracing algorithm returns whether a valid intersecting primitive was discovered and additional information about that intersection (e.g., primitive ID, normal, 3D point, and depth are all integers).
Overall, the quantum ray tracing algorithm requires $qb + 2\times cb + 5$ qubits to run and the depth of the circuit is $3 \times qb + \mathcal{O}( \lfloor  \frac{\pi}{4} \sqrt{2^{qb + 2\times cb + 5}} \rfloor)$. Therefore, when the quantum computers of edge servers can support this amount of qubits and execute the circuits, mobile devices can offload ray tracing applications to edge servers for remote execution and quantum acceleration.

\section{The Learning-based Algorithm Design}
To solve the computation offloading problem via RL algorithms, we first transform the problem into the POMDP, where each mobile device u is an agent interacting with the MEQC environment independently. Then, we design a hybrid discrete-continuous multi-agent RL algorithm for learning the optimal offloading and partitioning strategy.
  
  \begin{figure*}[t]
  	\centering
  	\includegraphics[width=1\linewidth]{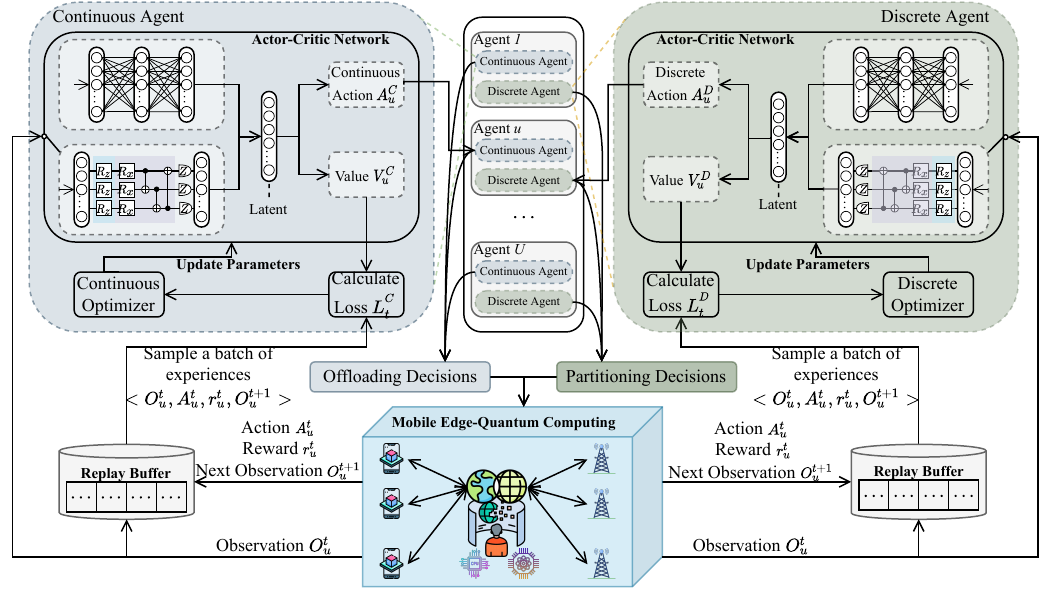}
  	\caption{The proposed hybrid discrete-continuous multi-agent reinforcement algorithms. In the algorithm, each learning agent consists of the continuous agent and the discrete agent, whose actor-critic networks can be parameterized by classic/quantum neural networks.}
  	\label{fig:algorithm}
  \end{figure*}
\subsection{POMDP of Quantum Computation Offloading for Mobile Edge-Quantum Computing}
  In MEQC, multiple users in MEQC attempt to offload computation tasks to edge servers via wireless connections. The computation offloading, including classical computing and quantum computing applications, can be modeled as a multi-agent RL problem,
  where each mobile device trains a deep RL agent to make offloading and partitioning decisions by interacting with the MEQC environment. 
  
  For deep RL algorithms to solve the computation offloading problem, the computing process has to follow a POMDP, which can be represented by a 7-tuple $(S, {O}, {A}, {R}, P, {\Omega}, \gamma)$ consisting of the state space $S$, the observation spaces $O$, the action spaces $A$, the reward functions ${R}$, the transition probabilities $P$, the observation probabilities $\Omega$, and the discount factor $\gamma$. In the setting of multi-agent RL, each mobile device $u$ acts as a learning agent to continuously explore the environment, whose state is denoted as $S^t$ at time slot $t$, and improve its policy $\pi_u$.
\subsubsection{Observation Space} 
  Based on the observation probabilities, we first define the observation space $O_u^t=\Omega^t_u(S^t)$ of mobile device $u$ from the state $S^t$ at time slot $t$ as a union of the local computation conditions $L_u^t$, edge computation conditions $E_u^t$, and wireless connection conditions $W^t_u$, which can be defined as
  \begin{equation}
      O^t_u \triangleq \left\{L_u^t, E_u^t, W^t_u\right\}.
  \end{equation}
  Specifically, the local computation conditions $L_u^t \triangleq [f_u^l, s_u, n_u, Q^L_u, D^L_u]$ of user $u$ consists of the local computational capacity $f_u^l$, the data size $s_u$ of the raw computation task, the required CPU cycles $n_u$ per data size, and the width $ Q^L_u$ and depth $D^L_u$ of the compiled circuit of the task.
  The edge computation condition $E_u^t \triangleq [f_u^E, Q_u^E, k_1,\ldots, k_E]$ observed by user $u$ includes the classic computing capacity $f_u^E$, quantum computing capacity $Q_u^E$, and error correction’s concatenation levels $k_1,\ldots, k_E$ of edge servers. Finally, the wireless connection condition $W^t_u \triangleq [p_u, g_u(1), \ldots, g_u(E)]$ consists of the transmit power of user $u$ and channel gains among user $u$ and edge servers.
  
  
\subsubsection{Action Space}
  Each mobile device $u$ as a learning agent needs to maintain a hybrid discrete-continuous action space, which is denoted as 
  \begin{equation}
      A^t_u \triangleq A^C_u \bigcup A^D_u,
  \end{equation}
  where the discrete action space $A^{ C}_u\triangleq \{a_u\}$ is for the offloading decision and the continuous action space $A^{ D}_u\triangleq \{\phi_u\}$ is for the partitioning decision.
  Here, $a_u$ is the discrete action to determine the server that user $u$ should connect to, and $\phi_u$ is the continuous action to indicate the portion of the task that user $u$ should process locally. Let $A^t = [A^t_1, \ldots, A^t_U]$ be the joint action of all agents.
\subsubsection{Reward Function}
After the state transition from the current environment state to the next state, each learning agent of mobile device $u$ can gain rewards $r_u(S^t, A^t) = -C(\boldsymbol{a}, \phi)$ w.r.t. current state and actions.
\subsection{Multi-agent Policy Evaluation and Improvement}

We first present the hybrid discrete-continuous policies of learning agents. Then, we propose the hybrid policy iteration including policy evaluation and policy improvement. Finally, we introduce the hybrid quantum policies which are based on trainable VQC.

Each learning agent $u$ maintains the discrete actor-critic network $(\pi^\emph{D}_u, V^{\pi^\emph{D}_u})$ and the continuous actor-critic network $(\pi^\emph{C}_u, V^{\pi^\emph{C}_u}_u)$ for determining discrete and continuous actions, respectively~{\cite{neunert2020continuous}}. In multi-agent policy evaluation and improvement, each learning agent first collects experiences during the interaction with the environment into its replay buffer. Then, the performance of the current strategy is evaluated using the critic networks with general advantage estimation. Finally, based on the evaluation of the critic networks, the policy networks are improved via gradient ascent w.r.t. the learning rate, while the critic networks are updated via gradient descent w.r.t. the learning rate. Let $\mathbb{E}_{\pi} (\cdot)$ denote the expected value of a random variable given that the agent follows policy $\pi$ and $\gamma \in [0, 1]$ which is the reward discount factor used to reduce the weights as the time step increases. Finally, the expected long-term value $V^D$ and $V^C$ are maximized, and thus the sustainability of the MEQC system is optimized.

\subsubsection{Hybrid Discrete-Continuous Policies}

In quantum computation offloading, the offloading decisions and the partitioning decisions are independent and can be performed simultaneously. Let $\vartheta_u$ and $\theta_u$ denote the trainable parameters in the discrete actor-critic network and the continuous actor-critic network of user $u$, respectively. The stochastic policy $\pi_u^{\tmop{hyb}} (A_u | O_u)$ of user $u$ to represent the hybrid discrete-continuous policy of agent $u$, which can be represented as
\begin{equation}
\begin{aligned}
    \pi_u^{\tmop{hyb}} (A_u | O_u) &=  \pi_{\theta_u }^D (A_u^D | O_u ) \pi_{\theta_u }^C (A_u^C | O_u) \\&= \prod_{A^i \in A^D} \pi_{\vartheta_u }^D
(A^i | O_u ) \prod_{A^i \in A^C}
\pi_{\theta_u }^C (A^i | O_u ),
\end{aligned}
\end{equation}
where $A^i$ denotes either discrete and continuous random variables, $A^C$ and $A^D$ are the sub-sets of action dimensions with continuous variables for partitioning decisions and discrete variables for offloading decisions, respectively.

Similar to Q-learning, the output of discrete policies are parameterized by state-dependent probabilities $\alpha_{\vartheta_u} (O_u)$. Therefore, the discrete policy $\pi_{\vartheta_u }^D$ follows the categorial distribution over $N$ discrete actions which can be represented as
\begin{equation}
\pi_{\vartheta_u }^D (A^i | O_u ) = \tmop{Cat}^i (\alpha_{\vartheta_u }
(O_u)).
\end{equation}
Meanwhile, for the stochastic continuous policies, the variables of the continuous policy of agent $u$ can be represented as the form of normal distribution, i.e.,
\begin{equation}
\pi_{\theta_u }^C (A^i | O_u ) = \mathcal{N} \left(\mu_{i, \theta_u} (O_u),
\sigma_{i, \theta_u}^2 (O_u) \right).
\end{equation}
Typically, these distributions of actions $\mu_{i, \theta} (O_u), \sigma_{i,
	\theta}^2 (O_u)$, and $\alpha_{\vartheta_u } (O_u)$ are output by the continuous
actor-critic network and the discrete actor-critic network, respectively. To train the hybrid policies to perform sustainable offloading and partitioning decisions in quantum computation offloading, the policies are first evaluated by a value function and then improved by stochastic gradient ascent.

\subsubsection{Hybrid Policy Evaluation in Multi-agent RL}

In the traditional setting of RL, the objective of agents is to learn the policy $\pi  (\cdot | O_u )$ to maximize the expected long-term return,
\begin{equation}
\mathbb{E}_{\pi \sim P} \left[ \sum_{t = 0}^{\infty} \gamma^t r_u (O_u^t,
A^t_u) \right],
\end{equation}
where $\gamma \in (0, 1]$ is a discount factor for allowing the sum of discounted rewards to converge over an infinite time horizon.

To evaluate the policy $\pi_{u}^{\tmop{hyb}}$ of user $u$, the learning agent tries to learn an action-value function, i.e., Q-function, to approximate the return according to a policy $\pi$, which can be represented as
\begin{equation}
\begin{aligned}
    Q^{\pi_u^{\tmop{hyb}} } (O_u, A_u) = \mathbb{E }_{\pi_u^{\tmop{hyb}} = (\pi^D_{\vartheta_u}, \pi^C_{\theta_u})} &\Bigg[ \sum_{t = 0}^{\infty}
\gamma^t r_u \left( O^t_u, A^t_u \right) \\&| O^0_u = O_u, A^0_u = A_u \Bigg].
\end{aligned}
\end{equation}
In the recursive expression, the discrete action-value function can be expressed as
\begin{equation}
\begin{aligned}
    Q^{\pi_{\vartheta_u}^D } (O^t_u, A^t_u) = \mathbb{E }_{O^t_u , A^t_u, O^{t + 1}_u \sim P}&
\big[ r_u (O^t_u, A^t_u) \\&+ \gamma V^{\pi_{\vartheta_u}^D}  (O^{t + 1}_u) \big],
\end{aligned}
\end{equation}
where the discrete state-value function $V^{\pi_{\vartheta_u}^D } (O^{t + 1}_u)$ is the expected action-value function, i.e., $V^{\pi_{\vartheta_u} } (O^{t + 1}_u) = \mathbb{E}_{\pi_{\vartheta_u}^D} \left[
Q^{\pi_{\vartheta_u}^D } ( O^{t + 1}_u , u^{t + 1}) \right]$, which
can indicate the expected return starting from observation $O^{t + 1}_u$ according to the
actions $u^{t + 1} \sim \pi_{\theta_{\vartheta_u}^D} (\cdot | O^{t + 1}_u)$ outputted
by the policy $\pi_{\vartheta_u}$ of user $u$. Meanwhile, the continuous action-value function can be expressed as 
\begin{equation}
\begin{aligned}
        Q^{\pi_{\theta_u}^C } (O^t_u, A^t_u) = \mathbb{E }_{O^t_u , A^t_u, O^{t + 1}_u \sim P}&
\big[ r_u (O^t_u, A^t_u) \\&+ \gamma V^{\pi_{\theta_u}^C}  (O^{t + 1}_u) \big],
\end{aligned}
\end{equation}
where the continuous state-value function $V^{\pi_{\theta_u}^C } (O^{t + 1}_u)$ is the expected action-value function, i.e., $V^{\pi_{\theta_u}^C } (O^{t + 1}_u) = \mathbb{E}_{\pi_{\theta_u}^C} \left[
Q^{\pi_{\theta_u}^D } ( O^{t + 1}_u , u^{t + 1}) \right]$.
To evaluate the relative advantage of that hybrid policy $\pi_{\theta_u}^{\tmop{hyb}}$, we
then define the advantage function as the difference between the action-value
function and state-value function as
\begin{equation}
A^{\pi_{u}^{\tmop{hyb}} } (O_u, A_u) = Q^{\pi_{u}^{\tmop{hyb}} } (O_u, A_u) - V^{\pi_{u}^{\tmop{hyb}}
} (O_u).
\end{equation}
However, calculating the advantage function exactly is computationally expensive and requires full knowledge of the environment's dynamics. Therefore, we leverage the truncated version of the generalized advantage estimation to evaluate the improvement of current policy over a trajectory with $T$ time steps, which can be represented as
\begin{equation}
\hat{A}^{\pi_{u}^{\tmop{hyb}}} = \delta_t + (\gamma \lambda) \delta_{t
	+ 1} + \cdots + (\gamma \lambda)^{T - t + 1} \delta_{T - 1},
\end{equation}
where $\delta_t = r_u (O^t_u, A^t_u) + \gamma V^{\pi_{\vartheta_u}^D} (O^{t + 1}_u) -
V^{\pi_{\vartheta_u}^D} (O^t_u) + \gamma V^{\pi_{\theta_u}^C} (O^{t + 1}_u) -
V^{\pi_{\theta_u}^C} (O^t_u)$ and $\lambda$ is a smoothing parameter to achieve variance reduction during training.

\subsubsection{Hybrid Policy Improvement in Multi-agent Learning}

In the M2RL algorithm, we leverage the clip-based proximal policy optimization method, and update the discrete and continuous policies via
\begin{equation}
\vartheta^{t + 1}_u = \arg \max_{\vartheta} \mathbb{E}_{O^t_u, A^t_u \sim
	\pi^{\tmop{hyb}}_{u} } [L^{D}_t (O^t_u, A^t_u, \vartheta_u)],
\end{equation}
and 
\begin{equation}
\theta^{t + 1}_u = \arg \max_{\theta} \mathbb{E}_{O^t_u, A^t_u \sim
	\pi^{\tmop{hyb}}_{u} } [L^{C}_t (O^t_u, A^t_u, \theta_u)],
\end{equation}
respectively. In practice, we take multiple steps of (usually minibatch) SGD to optimize the policy for maximizing the objective. Respectively, the discrete agent loss and continuous agent loss are given by
\begin{equation}
L^{D}_t (O^t_u, A^t_u, \vartheta_u) = L^{^{\tmop{D-clip}}}_t (\vartheta_u^t) -
L^{\tmop{D-VF}}_t (\vartheta_u^t) + c_2 S [\pi_{\vartheta_u}^D] (O^t_u),
\end{equation}
and 
\begin{equation}
L^{C}_t (O^t_u, A^t_u, \theta_u) = L^{^{\tmop{C-clip}}}_t (\theta_u^t) -
L^{\tmop{C-VF}}_t (\theta_u^t) + c_2 S [\pi_{\theta_u}^C] (O^t_u),
\end{equation}
where
\begin{equation}
L^{^{\tmop{D-clip}}}_t (\vartheta_u^t) = \min \left( \frac{\pi^D_{\vartheta_u^t} (O^t_u, A^t_u) }{\pi^D_{\vartheta_u^{\tmop{old}}} (O^t_u, A^t_u)} 
\hat{A}^{\pi_{\vartheta_u}^D} , g \left( {\epsilon},
\hat{A}^{\pi_{\vartheta_u}^D}  \right) \right),
\end{equation}
\begin{equation}
L^{^{\tmop{C-clip}}}_t (\theta_u^t) = \min \left( \frac{\pi^C_{\theta_u^t} (O^t_u, A^t_u) }{\pi^C_{\theta_u^{\tmop{old}}} (O^t_u, A^t_u)} 
\hat{A}^{\pi^C_{\theta_u}}, g \left( {\epsilon},
\hat{A}^{\pi^C_{\theta_u}}  \right) \right),
\end{equation}
\begin{equation}
g \left( {\epsilon}, \hat{A} \right) =
\left\{\begin{array}{ll}
(1 +{\epsilon}) \hat{A}, &
\hat{A} \geq 0,\\
(1 -{\epsilon}) \hat{A}, &
\hat{A} < 0,
\end{array}\right.
\end{equation}
$c_1$, $c_2$ are coefficient, $L^{\tmop{D-VF}}_t (\vartheta_u) = \left( V
^{\pi_{\vartheta_u}^D} (O^t_u) - V^{\tmop{targ}}_t \right)^2$ and $L^{\tmop{C-VF}}_t (\theta_u) = \left( V
^{\pi_{\theta_u}^C} (O^t_u) - V^{\tmop{targ}}_t \right)^2$ are squared-error losses
between the current  discrete/continuous values and the target value $V^{\tmop{targ}}_t$, and $S[\cdot]$ denotes an entropy function.

\subsection{VQC-based Quantum Hybrid Policies}

In addition, to parameterize the actor-critic networks using classical neural
networks, such as MLP, VQCs can also be
leveraged to parameterize the policies and value functions of deep RL agents. VQCs
are a type of quantum circuit that can be used as function approximators in a
classical RL setting. In the application of VQCs in
quantum RL, VQCs are used as a quantum version of MLP with
their adjustable parameters. Typically, each layer of VQCs consists of three
blocks, i.e., data-encoding circuit blocks $\widehat{S } (s )$, parameterized
circuit blocks $\hat{U} (\theta)$, and non-parametrized circuit blocks
$\hat{V}$. Depending on the input observation of user $u$, data-encoding
circuit blocks are responsible for translating classical data into quantum
states, which are then used as input for quantum machine learning algorithms.
Including a set of trainable parameters, the parameterized circuit blocks can
be adjusted by using optimization techniques such as stochastic gradient
descent. The policy improvement process is iterative, and it involves
computing the gradient of the cost function with respect to the parameters and
updating the parameters accordingly. The number of iterations required to find
the sustainable parameters depends on the complexity of the problem and the
resources available. Finally, the non-parameterized circuit blocks are leveraged to entangle qubits, such as the CNOT gate used for entangling qubits. Overall, the computing process of VQC-based quantum hybrid policy of user $u$ can be represented as a unitary
\begin{equation}
\hat{\mathcal{U}}_{\theta_u, \vartheta_u} (O^t_u) = \hat{V} \hat{U} (\theta_u, \vartheta_u) \widehat{S }
\left( {O^t_u}  \right) .
\end{equation}
With multiple runs of the circuit, the prediction from such a model is then evaluated as the expectation value of an observable $M$, with respect to the the final state of the quantum circuit, which can be presented as
\begin{equation}
\pi_{\theta_u, \vartheta_u }^{\tmop{hyb}} (A^t_u | O^t_u ) = \langle \psi_0 |
\hat{\mathcal{U}}_{\theta_u, \vartheta_u} (O^t_u)^{\dag} M \hat{\mathcal{U}}_{\theta_u, \vartheta_u}
(O^t_u) | \psi_0 \rangle,
\end{equation}
where $| \psi_0 \rangle$ is some initial state of the quantum system.

During the inference stage of the proposed hybrid RL algorithm, the computational complexity is $\mathcal{O}(2UH)$, where $U$ is the number of learning agents and $H$ is the computation of each actor-critic network. For the agent parameterized by the deep neural network, the computation can be represented as $H = H^D$, while for the agent parameterized by the quantum neural network, it can be represented as $H = H^Q$. Generally, $H^D \gg H^Q$ is attributed to the computing acceleration of quantum computing. This phenomenon can also be referred to as more sustainable in the inference stage.

		  \begin{figure*}[t]
	\centering
	\begin{minipage}[t]{0.33\linewidth}
		\centering
		\includegraphics[width=1\linewidth]{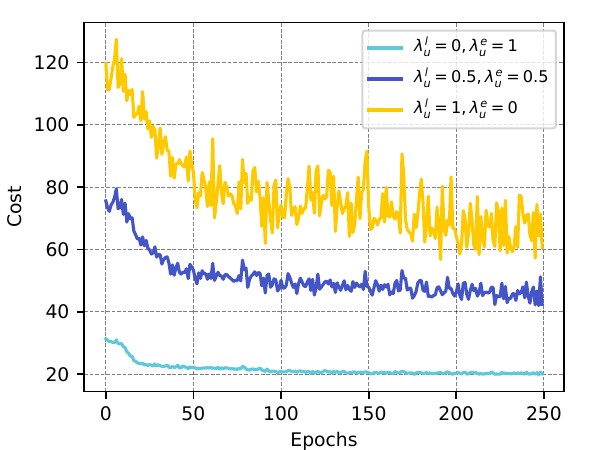}
		\caption{System cost v.s. training epochs, \textit{U}=10 and \textit{E}=10.}
		\label{fig:con10}
	\end{minipage}%
	\begin{minipage}[t]{0.33\linewidth}
		\centering
		\includegraphics[width=1\linewidth]{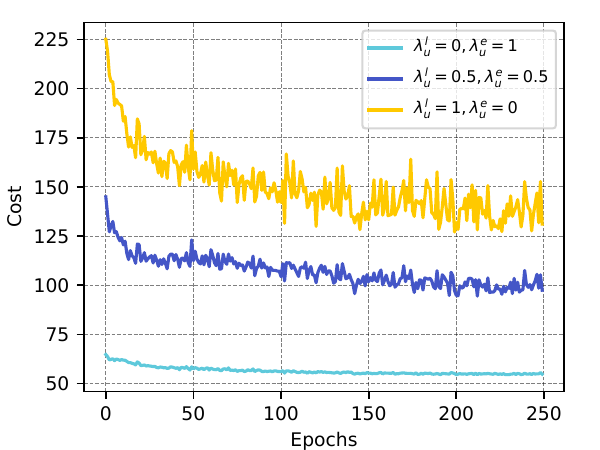}
		\caption{System cost v.s. training epochs, \textit{U}=20 and \textit{E}=20.}
		\label{fig:con20}
	\end{minipage}
	\begin{minipage}[t]{0.33\linewidth}
		\centering
		\includegraphics[width=1\linewidth]{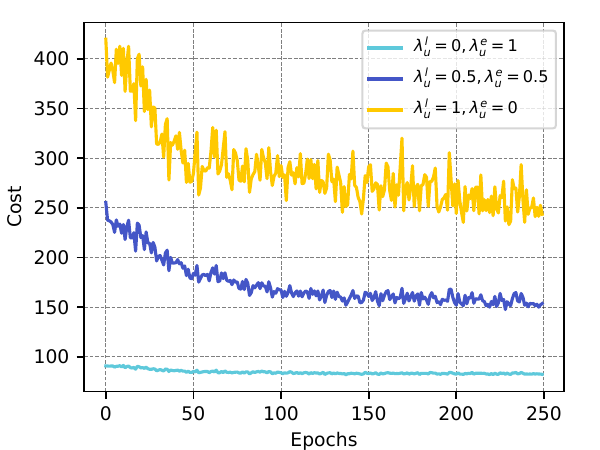}
		\caption{System cost v.s. training epochs, \textit{U}=30 and \textit{E}=30.}
		\label{fig:con30}
	\end{minipage}
\end{figure*}
		\section{Experimental Results}\label{sec:exp}
  In this section, we first present the parameter settings in the MEQC system. Then, we provide a convergence analysis of the proposed HMADRL algorithm and demonstrate the performance comparison between the proposed algorithm and baselines. Finally, we show the sustainability of the proposed RL algorithm with learning agents parameterized by quantum neural networks.
\subsection{Parameter Settings}

  In the simulation experiment, we first consider a MEQC system with 10 mobile devices and 10 edge servers. 
  For the communication model, the channel gain of each user is randomly assigned from the set [4, 8], and the transmission power is allocated from [0.01, 0.2] mWatts. The bandwidth owned by each server is set to 20 MHz. For the classical computation model, the local computational capacity is randomly assigned from the set \{1, 2, 3\} GHz, and the edge computation capacity is randomly assigned from the set \{10, 15, 20\} GHz. The chip coefficient is assigned to $\gamma=10^{-11}$ for the energy consumption per CPU cycle according to the measurement method as given in \cite{chen2015efficient}. For the quantum computation model, the number of physical qubits at each edge server is randomly assigned from [1000, 5000]~\cite{resch2021benchmarking}, the error correction's concatenation level is randomly selected from [1, 2, 3], the qubit operation temperature is set to 0.1 K, the signal generation temperature is set to 300 K, and the attenuation is set to 40 dB. 
  In line with~\cite{fellous2022optimizing}, the typical qubit frequency is set to 6 GHz, the 1qb gate latency is set to 25 ns, the 2qb gate latency is set to 100 ns, the measurement latency is set to 100 ns, the number of refrigeration stages is set to 5, the threshold for error correction is set to $2\times 10^{-4}$, the heat produced by signal generation \& readout is set to 10 $\mu$W, the heat produced at 4K by params is set to 10 nW, and the heat produced at 70K by HEMT amps is set to 50 $\mu$W. For each logical qubit with error correction's concatenation $k$, the number of required physical qubits is $(91)^k$, the number of physical 1qb gates is $\frac{28}{185}(64)^{k}$, the number of physical 2qb gates is $\frac{64}{185}(64)^{k}$, and the number of physical measurement gates is $\frac{28}{185}(64)^{k}$. More details of the calculation of physical qubits and gates in error correction are listed in Appendix~\ref{app:error}.
  
  In this paper, we focus on ray-tracing rendering tasks, where the coordinate is set to 16$\times$16$\times$16, the resolution of each frame is set to 128$\times$128, the number of frames is randomly assigned from [1024, 10240], the number of primitives is randomly assigned from \{3, 4, 5, 6, 7, 8, 9\}, the number of rays of each primitive is set to 3. Therefore, the data size of each classical computation task is randomly assigned from [160, 1600] MB and the number of required CPU cycles is randomly assigned from 3 $\times$ \{$2^3,\ldots,2^9$\} cycles/byte. In addition, when the classical computation task is converted to a quantum computation task, the required numbers of qubits are $\{20,\ldots, 26\}$ and the required circuit depths are \{813,\ldots, 6560\}. For the weights of each user $u$ for both the latency and energy, they are set to 0.5 by default.

  The hyperparameters in the proposed deep RL algorithm are set as follows. The discrete and the continuous policies are parameterized by two-layer fully connected networks with 256 hidden units. The QNN includes one 64-unit input layer, one 64-unit output layer, and one VQC consisting of one AngleEmbedding layer, two 4-qubit BasicEntangler layers, and one PauliZ layer. The quantum circuit is implemented by pennylane~\cite{bergholm2018pennylane}. We train the classic algorithm for 250 epochs and the quantum algorithm for 125 epochs, where each epoch consists of 500 steps. After each epoch, the policies are updated twice with a batch size of 128, a discounting factor of 0.95, and a learning rate of 0.001. We perform our experiment with Python 3.8, PyTorch 1.12.1, CUDA 11.6, and cuQuantum 23.3.0.

    \begin{figure*}[t]
	\centering
        \begin{minipage}[t]{0.33\linewidth}
		\centering
		\includegraphics[width=1\linewidth]{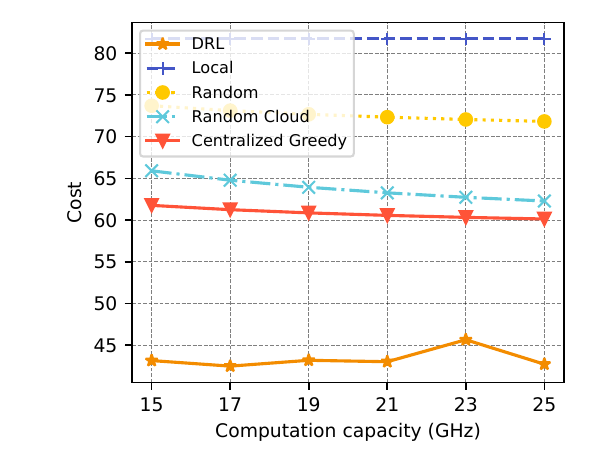}
		\caption{System cost v.s. classical computation capacity.}
		\label{fig:classical}
	\end{minipage}%
	\begin{minipage}[t]{0.33\linewidth}
		\centering
		\includegraphics[width=1\linewidth]{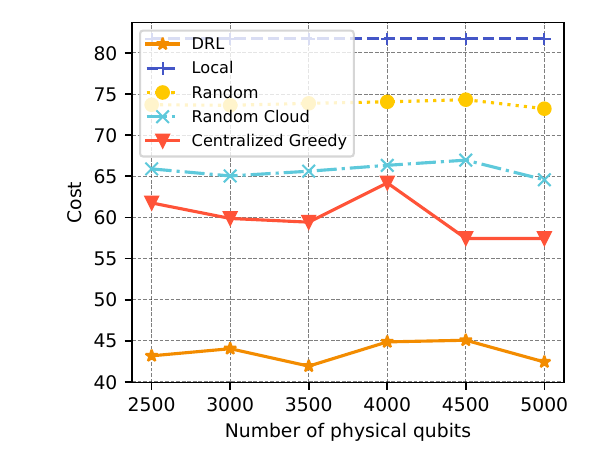}
		\caption{System cost v.s. quantum computation capacity.}
		\label{fig:quantum}
	\end{minipage}
	\begin{minipage}[t]{0.33\linewidth}
		\centering
		\includegraphics[width=1\linewidth]{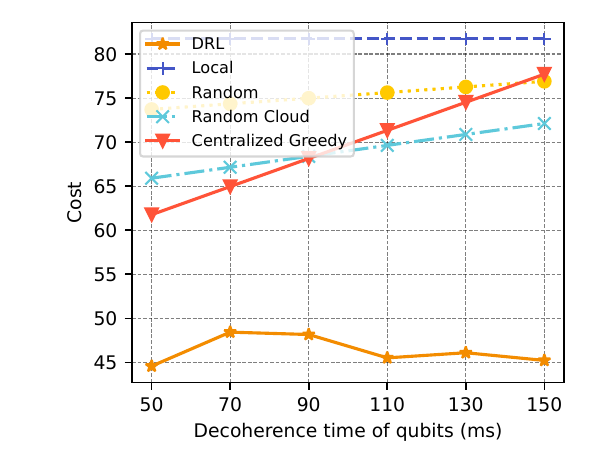}
		\caption{System cost v.s. qubit quality.}
		\label{fig:qtime}
	\end{minipage}
\end{figure*}
  
  \subsection{Convergence Analysis}

  First, we analyze the convergence of the proposed hybrid discrete-continuous multi-agent deep RL. In Fig.~\ref{fig:con10}, we show the performance of the proposed algorithm in achieving convergence in user scenarios with different preferences. For the more energy-aware scenarios, the proposed algorithm takes about 50 epochs to achieve convergence performance. On the other hand, for the more delay-aware scenario, the proposed algorithm takes about 100 epochs to reach the convergence performance, which is similar to that of the scenario where the delay and energy are equally weighted. From Figs. \ref{fig:con20} and \ref{fig:con30}, we can observe that the proposed algorithm can converge to the stable offloading and partitioning strategies at around 200 epochs when there are 20 users/servers and 30 users in the MEQC system, respectively. 
  
  \subsection{Performance Comparison}

  Then, we evaluate the proposed system model and the proposed deep RL algorithm under different system settings. We use local computing, random offloading, random edge partitioning, and centralized greedy schemes as the baseline solutions for performance comparison. In Fig. \ref{fig:classical}, for all the algorithms except the local execution algorithm, the cost of the algorithm decreases as the computation capacity of the edge server increases. However, as we can observe from Fig. \ref{fig:quantum}, the increase in the number of qubits will not significantly affect the consumption of computation offloading in MEQC. The reason is that the number of qubits at edge servers is a hard-cutting success probability to determine whether the quantum advantage can be brought to mobile devices. Finally, we increase the quality of qubits (i.e., decoherence time) in quantum computers and illustrate the results in Fig.~\ref{fig:qtime}. The costs of quantum computing increase dramatically as the quality of qubits increases. Nevertheless, the proposed deep RL algorithm leads to choosing the most economical offloading and partitioning strategy without the impacts of increasing the energy consumption from quantum computing. Overall, the proposed algorithm can reduce at least 30\% of the cost compared with existing baselines

  
		\section{Conclusions}
		In this paper, we introduced mobile edge quantum computing (MEQC) for the Quantum Internet and formulated the hybrid classic-quantum computation offloading problem as mixed-integer programming via non-convex and nonlinear objectives. To make decentralized offloading and partitioning decisions for mobile devices, we transformed the problem into the POMDP where users act as learning agents. We proposed a hybrid discrete-continuous MARL to learn the sustainable offloading and partitioning strategies. Numerical results showed that the proposed algorithms can improve system performance in terms of convergence rate and system cost compared with the existing baselines under different system settings. For future work, we will explore collaborative mobile edge-quantum computing, allowing mobile users to choose multiple BSs for executing their computation tasks collaboratively.
		
		
		%

        		\appendix
  \subsection{The Noise Model of Scalable Quantum Computers}\label{app:noise}
  In quantum computing, noise is defined as any undesired interference between the quantum system and its surrounding environment, which potentially induces computational errors. A noise model for a scalable quantum computer provides a mathematical depiction of different noise sources that may influence the system's qubits and gates. This model is instrumental in comprehending quantum system behavior and formulating strategies to minimize noise interference. Components like decoherence, gate errors, and measurement errors are considered in the creation of the noise model.
  We take the common concept that stages should have equal attenuation and be regularly spaced in orders of magnitude of temperature between $T_{\tmop{gen}}$ and $T_{\tmop{qb}}$~\cite{martin2022energy}. In other words, if we want a total attenuation of $A$, we take
\begin{equation}
  A_i = A^{1 / (K - 1)}, \quad T_i = T_{\tmop{qb}} \left(
  \frac{T_{\tmop{gen}}}{T_{\tmop{qb}}} \right)^{(i - 1) / (K - 1)},
\end{equation}
for $K$ stages of cooling of quantum computers. The cooling stages of a quantum computer are determined based on the required temperature for the qubits to maintain their quantum coherence. The cooling process involves several stages, each using a different cooling method, such as a dilution refrigerator or a pulse-tube refrigerator. The final stage of cooling is typically achieved using a helium-3 refrigerator or a dilution refrigerator. The number of cooling stages required depends on the specific quantum computing system and the desired operating temperature.

We consider the thermal photon contribution to the noise to be reduced to an acceptable level by a chain of attenuators on the ingoing microwave line. These attenuators are kept cold by cryogenics, and hence they thermalize the signal to come down the line from hotter temperatures, reducing the population of thermal photons. For a chain of $K$ cooling stages with $K - 1$ attenuators (e.g., $K = 5$), the error probability of a physical qubit is~\cite{fellous2022optimizing}
\begin{equation}
  \epsilon_{\tmop{err}} = \frac{\gamma \tau_{\tmop{step}}}{2} \left( \frac{1}{2} + n
  (T_1) + \sum^{K - 1}_{i = 1} \frac{n (T_{i + 1}) - n (T_i)}{\tilde{A}_i} 
  \right),\label{error}
\end{equation}
where $T_1 = T_{\tmop{qb}}$, and $n(T)  = (\exp [\hbar \omega / k_B T] - 1)^{- 1}$
is the Bose-Einstein function at the qubit frequency. In Eq. \eqref{error}, we observe that the noise can always be reduced by increasing the attenuation, which results in higher power consumption. In quantum computers, attenuation is a technique used to reduce the amplitude of a signal, which in turn reduces the noise in a quantum system. Increasing the attenuation can further reduce the noise, but it also results in higher power consumption, which is a crucial parameter in the optimization of the system. This trade-off between noise reduction and power consumption needs to be carefully considered in the design and optimization of quantum computing systems.

		\subsection{The Energy Consumption Model of Quantum Computers}\label{app:energy}
The resource cost is defined as the power $P_{\pi}$ consumed to bring the
qubit from $|0\rangle$ to $|1\rangle$, which can be defined as $
    P_{\pi} = \frac{\hbar \omega_0 \pi^2}{4 \gamma \tau^2_{1}},$
where $\omega_0$ is the transition frequency and $\gamma^{-1}$ is the spontaneous emission rate, i.e., the decoherence time, depending on specific qubit technology.
The power consumption per physical 2qb gate averaged over the timestep of the quantum computer can be defined as
  $P_{2} = P_{\pi} \sum_{i = 1}^K \frac{T_{\tmop{gen}} - T_i}{T_i}
  (\tilde{A}_i - \tilde{A}_{i - 1})$,
where $T_{\tmop{gen}}$ is the generation temperature, $T_i$ is the intermediate
temperature at stage $i$, and $\tilde{A}_i = A_i\times \cdots \times A_2 \times A_1$ is the total
attenuation between $T_i$ and the qubits. Moreover, the power consumption per physical 1qb gate can be defined as
  $P_{1} = \frac{\tau_{1}}{\tau_{\tmop{step}}} P_{2}$,
where $\tau_{1}$ is the 1qb gate latency and $\tau_{\tmop{step}}$ is the timestep of quantum computers.
Finally, the power consumption per physical qubit is
\begin{equation}
    P_{\mathrm{Q}} = \frac{T_{\text{ext}}}{T_{\text{gen}}} 
  \dot{q}_{\text{gen}} + \frac{T_{\text{ext}}}{T_{\text{hemt}}} 
  \dot{q}_{\text{hemt}} + \frac{T_{\text{ext}}}{T_{\text{para}}} 
  \dot{q}_{\text{para}},
\end{equation}
where $\dot{q}_{\text{gen}}, \dot{q}_{\text{hemt}},$ and $ \dot{q}_{\text{para}}$ are heat produced at $T_{\text{ext}} = T_{\text{gen}}, T_{\text{hemt}}=70 K,$ and $T_{\text{para}}=4 K$, respectively.

\subsection{Calculating Qubits and Gates for Levels of Concatenated Error Correction}\label{app:error}
		
 		In the experimental part of this paper, we consider fault-tolerant quantum computing built from concatenating a 7-qubit code~\cite{fellous2022optimizing}. The reason is that the 7-qubit code is a well-studied scheme with fairly complete and well-documented analyses. Therefore, any resource requirements are not overlooked during the discussion of mobile-edge quantum computing. 
		
 		For fault-tolerant Clifford gates, each level of error correction (i.e., one concatenation level) replaces one logical qubit by 7 data qubits and uses 28 ancilla qubits to detect errors. During the error correction for a given gate, ancillas must be prepared for the next two gates, which takes three-time steps in the data qubits operations. Therefore, each additional level of concatenation in the error correction involves replacing one qubit with 91 qubits (7 data qubits and 3 $\times$ 28 ancillas). For a $k$-level error correction, the number of physical qubits is
 		\begin{equation}
 		    Q = (91)^k Q_L,
 		\end{equation}
 		where $Q_L$ is the number of logical qubits. In addition, this number is independent of the type of logical Clifford gates that is implemented on the logical qubits.
		
 		Then, with $k$ concatenation levels, the number of physical gates in parallel is related to the number of logical gates in parallel, by 
 		\begin{equation}
 		    \left(\begin{array}{c}
 N_{2 \mathrm{qb}} \\
 N_{1 \mathrm{qb}} \\
 N_{\mathrm{Id}} \\
 N_{\text {meas }}
 \end{array}\right)=A^{k}\left(\begin{array}{c}
 N_{2 \mathrm{qb} ; \mathrm{L}} \\
 N_{1 \mathrm{qb} ; \mathrm{L}} \\
 N_{\mathrm{Id} ; \mathrm{L}} \\
 N_{\text {meas } ; \mathrm{L}}
 \end{array}\right)
 		\end{equation}
 		with
 		\begin{equation}
 		    A=\frac{1}{3}\left(\begin{array}{cccc}
 135 & 64 & 64 & 0 \\
 56 & 35 & 28 & 0 \\
 58 & 29 & 36 & 0 \\
 56 & 28 & 28 & 7
 \end{array}\right),
 		\end{equation}
 		where the elements of $A$ is defined in~\cite{fellous2022optimizing}. The prefactor of 1/3 in $A$ is because it takes three times of timesteps to perform each logic gate. 
 		Based on $A$, the number of logic gates can be approximated as
 		\begin{equation}
 		    \begin{aligned}
 N_{2 \mathrm{qb}} & \simeq \frac{64}{185}(64)^{k} Q_{\mathrm{L}}, \quad N_{1 \mathrm{qb}} \simeq \frac{28}{185}(64)^{k} Q_{\mathrm{L}}, \\
 N_{\mathrm{id}} & \simeq \frac{29}{185}(64)^{k} Q_{\mathrm{L}}, \quad N_{\text {meas }} \simeq \frac{28}{185}(64)^{k} Q_{\mathrm{L}},
 \end{aligned}
 		\end{equation}
 		where the $\simeq$ indicates the approximations and assumptions made in the previous paragraphs. For any quantum algorithm, the number of physical gates in parallel after $k\geq 1$ levels of concatenations is about the same. 
		
 		The calculation of Clifford gates can be efficiently simulated with classical computers, while the non-Clifford gates, i.e., T-gates, are required to perform an arbitrary quantum calculation. To simplify the presentation of the full-stack approach in this article, we made the brutal simplification to treating $T$-gates as requiring the same resources as Clifford gates~\cite{fellous2022optimizing}. Although the method used to make non-Clifford gates in a fault-tolerant manner is very different than for Clifford gates, the logical circuit contains few enough T-gates that have a negligible contribution to the total power consumption. For instance, a circuit with no T-gates, such as a quantum memory, whose job is to preserve an arbitrary quantum state. Another example could be algorithms in which the number of T-gates has been minimized to a tiny fraction of the total number of gates~\cite{kissinger2020reducing}.

    \bibliographystyle{IEEEtran}
		\bibliography{main}

	\end{document}